\documentclass[10pt,bezier]{article}
\usepackage{amsmath,amssymb,amsfonts, euscript}
\newcommand{\Real}{I\!\!R}
\newcommand{\R}{{\bf R}}
\newcommand{\ov}{\overline}
\newcommand{\N}{\nabla^\omega}
\newcommand{\g}{{\Bbb g}}
\newcommand{\B}{{\bf B}}
\newcommand{\te}{\textstyle}
\newcommand{\X}{\frak{X}}
\newcommand{\Oa}{\Omega^a}
\newcommand{\n}{\widehat \nabla}
\newtheorem{thm}{Theorem}[section]

\newtheorem{cor}[thm]{Corollary}

\newtheorem{defn}[thm]{Definition}

\newtheorem{prop}[thm]{Proposition}

\begin{document}
\noindent {\bf\Large Application of Lie algebroid structures to unification of Einstein and yang-mills field equations}\\

\noindent \\

\noindent {\bf N. Elyasi}\footnote{  E-mail:elyasi82@aut.ac.ir }\\
 Dep. of Math. \& Comp., Amirkabir
University of Technology, Tehran, Iran\\
\noindent {\bf  N. boroojerdian}\footnote{  E-mail: broojerd@aut.ac.ir }\\
 Dep. of Math. \& Comp., Amirkabir
University of Technology, Tehran, Iran
 \\

\noindent \\

\noindent {\bf  Abstract}$\quad$ Yang-mills field equations
describe new forces in the context of Lie groups and principle
bundles. It is of interest to know if the new forces and
gravitation can be described in the context of algebroids. This
work was intended as an attempt to answer last question. The basic
idea is to construct Einstein field equation in an algebroid
bundle associated to space-time manifold. This equation contains
Einstein and yang-mills field equations simultaneously. Also this
equation yields a new equation that can have interesting
experimental results.\\


%\checkmark
\section{Introduction}
 Einstein field equation describes gravitational forces, and yang-mills field equations describe other forces.
 Principle bundles on a space-time and principle connections are main apparatus for introducing and compromising  yang-mills
 theory and GR [1]. This method as in Kaluza-Klien theory [6],[9] make us to assume some extra dimension in space-time. Here we
 propose some other method that is capable of describing gravitation and new forces
 simultaneously and needs no extra dimension in space-time. The main idea of this method is enriching  tangent bundle of the
 space-time by a lie algebra  and make an algebroid structure.
   In our method, we need no extra dimension in space-time
 manifold, but we add some extra dimension to tangent bundle of
 space-time. Our approach is different from Kaluza-Klien
 theory,since first we have no extra dimension in space-time, second we use
 different mathematical structures, third our method is more
 general and contains results of yang-mills theory. Of course, our
 results is very near to that of Kaluza-Klien and yang-mills
 theory, but our method is different.

In [3] we have introduced an Lie algebroid bundle that is an
extension of tangent bundle of some space-time. This extension
can be expressed by
\[ \widehat {TM}=\sum_pT_pM\oplus\Real=TM\oplus\Real \]
In that structure we could unify Einstein and Maxwell equations.
In that method $\Real$ plays the role of a trivial Lie algebra. In
this paper we replace $\Real$ with  an arbitrary Lie algebra
$\g$. One of the physical interpretations of this work, is
replacing the unified Maxwell and Einstein field equations in [3]
with the unified yang-mills and Einstein field equations.

\section{Connection form and its curvature form}
In this section $M$ is a smooth manifold and $U\in\frak XM$ and
$\g$ is a Lie algebra and $\omega \in A^1(M,{\Bbb g})$. $\omega$
has some relations to connection forms on principle bundles and
may be used to define connection on some trivial vector bundles,
so we call it a $\g$-valued connection form on $M$(note this is
just a name and it doesn't mean that $\omega$ satisfies the
connection form conditions). Let $W$ be vector space and $\rho
:\g\times W \longrightarrow W \ ,\ \rho(h,v)=h.v$ be a Lie
algebra representation of $\g$ on $W$.

We can use $\omega$ to define a connection on the trivial vector
bundle $M\times W$ by the following equation.

\begin{equation}
X\in C^\infty(M,W),\quad \nabla^\omega_UX=U(X)+\omega(U).X
\end{equation}
By $U(X)$ we mean Lie derivation of $X$ with respect to $U$. As an
important example consider adjoint representation of $\g$ on
itself. If $\xi\in C^\infty (M,\g)$, then from $(1)$ the next
equation holds.
\[ \nabla^\omega_U\xi=U(\xi)+[\omega(U),\xi] \]
\begin{prop}
If $\xi\in C^\infty (M,\g)$ and $ X\in C^\infty(M,W)$ , then we
have
\begin{equation}
 \N_U(\xi .X)=(\N_U\xi).X+\xi.\N_UX
\end{equation}
\end{prop}
{\bf proof:} Since the action of $\g$ on $W$ is bilinear, then
 \[ U(\xi .X)=U(\xi).X+\xi .U(X) \]
Now the following computations show the result.
\begin{eqnarray*}
\N_U(\xi .X)&=& U(\xi .X)+\omega(U).(\xi .X)\\
 &=&U(\xi).X+\xi .U(X)+\omega(U).(\xi .X)-\xi.(\omega(U) .X)+\xi.(\omega(U) .X)\\
 &=&U(\xi).X+\xi .(U(X)+\omega(U) .X)+[\omega(U),\xi].X\\
 &=&(\N_U\xi).X+\xi.\N_UX \quad \blacksquare
\end{eqnarray*}
In the case of $M\times\g$, if $\xi,\eta \in C^\infty (M,\g)$ ,
then
\[
\N_U[\xi ,\eta]=[\N_U\xi,\eta]+[\xi,\N_U\eta]
\]
By direct computation we find curvature tensor of $\nabla^\omega$. For $U,V\in\frak XM$ and $X\in C^\infty(M,W)$
assuming $[U,V]=0$, we have
\begin{eqnarray*}
R^\omega (U,V)(X)&=&\N_U\N_VX-\N_V\N_UX \\
 &=&\N_U(V(X)+\omega(V).X)-\N_V(U(X)+\omega(U).X)\\
 &=&UV(X)+U(\omega(V).X)+\omega(U).V(X)+\omega(U).(\omega(V).X)\\
 &&-VU(X)-V(\omega(U).X)-\omega(V).U(X)-\omega(V).(\omega(U).X)\\
 &=&U(\omega(V)).X+\omega(V).U(X)+\omega(U).V(X)\\
 &&-V(\omega(U)).X-\omega(U).V(X)-\omega(V).U(X)+[\omega(U),(\omega(V)].X\\
 &=&(d\omega(U,V)+[\omega(U),\omega(V)]).X
\end{eqnarray*}
Define $\Omega\in A^2(M,\g)$ as follows.
\begin{equation}
2\Omega(U,V)=d\omega(U,V)+[\omega(U),\omega(V)]
\end{equation}
So,
\begin{equation}
R^\omega (U,V)(X)=2\Omega(U,V).X
\end{equation}
\\In the case of $M\times\g$, if $\xi\in C^\infty (M,\g)$,
 then $R^\omega (U,V)(\xi)=2[\Omega(U,V),\xi]$. We call $\Omega$ the
curvature form of $\omega$.

Assume $W$ has a positive definite inner product and $\g$ acts on
$W$ anti-symmetrically i.e.
\[
\forall h\in\g,\ \forall u,v\in W,\ \ \ <h.u,v>=-<u,h.v>
\]
For example, if a Lie group acts isometrically on $W$, then its
Lie algebra acts anti-symmetrically on $W$. Specially if a Lie
group has a bi-invariant metric, then the adjoint representation
of its Lie algebra on it self is  anti-symmetric.

By this assumption, $M\times W$ is a Riemannian vector bundle and
$\N$ is a Riemannian connection i.e.
\[
X,Y\in C^\infty (M,W),\ \ U<X,Y>=<\N_UX,Y>+<X,\N_UY>
\]

\section{Semi-Riemannian Lie algebroid $TM^\g$}
In this section $M$ is a semi-Riemannian manifold , $U\in\frak XM$
and $\g$ is a Lie algebra which has an inner product and its
adjoint representation action is anti-symmetric, and $\omega$ is
a $\g$- valued connection form on $M$.

Set $TM^\g=\cup_{p\in M}(T_pM\oplus\g)$. $TM^\g$ is a vector bundle and its sections has the form
$V+\xi$ in which $V\in\X M$ and $\xi\in C^\infty (M,\g)$.  $TM^\g$  has a natural  Lie algebroid
structure by the anchor map $\rho (V+\xi)=V$ and the following Lie bracket.
\begin{equation}
[U+\xi,V+\eta]=[U,V]+[\xi,\eta]+U(\eta)-V(\xi)
\end{equation}
Straightforward computations verify that $TM^\g$ is a Lie
algebroid. By inner product of $\g$ and metric of $M$ and
$\omega$, we can define a semi-Riemannian metric on $TM^\g$. As
in [3] we suggest that $TM$ is not orthogonal to $\g$, instead
some subbundle of $TM^\g$ isomorphic to $TM$, is orthogonal to
$\g$. This subbundle is denoted by $\overline{TM}$ and defined as
follows.

\begin{equation}
\overline{TM}=\{v+\omega(v)\ |\ v\in T_pM\ ,\ p\in M\}
\end{equation}
 We denote $v+\omega(v)$ by $\overline v$.For a
vector field $V\in\X M$, set \mbox{$\overline V=V+\omega(V)$}.
The meter of $TM^\g$ is defined as follows. For $U,V\in\X M,$ and
$\xi,\eta\in C^\infty (M,\g)$,
\begin{equation}
 <\overline U+\xi\ ,\ \overline V+\eta>= <U,V>_M
+<\xi,\eta>_\g
\end{equation}
Note that $\overline{TM}$ is not horizontal subbundle of some
connections, but $\overline{TM}$ is a subbundle of $TM^\g$ and is
complement to trivial subbundle $M\times\g$. In fact, by the
above definition $\overline{TM}$ is orthogonal subbundle of
$M\times\g$.
\\ Because of this definition it is better all
computations be done with respect to $\overline U$s and $\xi$ s.
For example, brackets of these sections of $TM^\g$ are computed as
follows.
\begin{eqnarray}
{[\overline U,\xi]}&=&\N_U\xi\\
{[\overline U,\overline V]}&=&\overline {[U,V]}+2\Omega(U,V)
\end{eqnarray}
Also, $\rho(\ov U)=U$. In foregoing  computations we need to use some other tensors equivalent to
the curvature form $\Omega$. $\Omega$ as an operator is $\Omega :\X M\times \X M\longrightarrow
C^\infty(M,\g)$. we define tensor  $\Oa :\X M\times C^\infty(M,\g)\longrightarrow \X M$ as
follows. For $U,V\in\X M$ and $\xi\in C^\infty (M,\g)$,
\begin{equation}
<\Oa(U,\xi),V>_M=<\Omega(U,V),\xi>_\g
\end{equation}
$\Oa(U,\xi)$ is anti-symmetric with respect to $U$.

\section{Levi-civita connection and curvature of $TM^\g$}
In this section, $M$ is a semi-Riemannian manifold and $U,V\in\X
M$, and $\g$ is a Lie algebra which has an inner product that the
action of its adjoint representation is anti-symmetric, and
$\xi,\eta\in C^\infty(M,\g)$, and $\omega$ is a $\g$- valued
connection form on $M$, and $TM^\g$ is the semi-Riemannian Lie
algebroid defined in the previous section.

Levi-civita connection of the semi-Riemannian Lie algebroid
$TM^\g$ is defined similar to semi-Riemannian manifolds [2]. We
denote this connection by $\n$ which is defined by the following
relation, if  $ \widehat U,\widehat V,\widehat W$ be arbitrary
sections of $TM^\g$ :
\[\begin{array}{rl}
 2<\n_{\widehat U}\widehat V,\widehat W> =& \rho (\widehat U)<\widehat V,\widehat W>+\rho (\widehat V)<\widehat W,\widehat U>
 -\rho (\widehat W)<\widehat U,\widehat V>\\
 & +<[\widehat U,\widehat V],\widehat W>-<[\widehat V,\widehat W],\widehat U>+<[\widehat W,\widehat U],\widehat V>
\end{array}\]
\begin{prop}
Levi-civita connection of the  $TM^\g$ satisfies the following relations.
\end{prop}
\begin{eqnarray}
\n_\xi\eta&=&{\te 1\over\te 2}[\xi,\eta] \\
\n_{\ov U}\xi&=&-\ov{\Oa(U,\xi)}+\N_U\xi \\
\n_\xi\ov U&=&-\ov{\Oa(U,\xi)} \\
\n_{\ov U}\ov V&=&\overline{\nabla_UV}+\Omega(U,V)
\end{eqnarray}
{\bf Proof:} Straightforward computations show these results. For
example we verify (12).
\begin{eqnarray*}
2<\n_{\ov U}\xi,\eta>&=& \rho (\ov U)<\xi,\eta>+\rho (\xi)<\eta,\ov U>
 -\rho (\eta)<\ov U,\xi>\\
 &&+<[\ov U,\xi],\eta>-<[\xi,\eta],\ov U>+<[\eta,\ov U],\xi>\\
 &=&U<\xi,\eta>+<\N_U\xi,\eta>-<\N_U\eta,\xi>\\
 &=&2<\N_U\xi,\eta>\\
 &&\\
2<\n_{\ov U}\xi,\ov V>&=&\rho (\ov U)<\xi,\ov V>+\rho (\xi)<\ov V,\ov U>
 -\rho (\ov V)<\ov U,\xi>\\
 &&+<[\ov U,\xi],\ov V>-<[\xi,\ov V],\ov U>+<[\ov V,\ov U],\xi>\\
 &=&<\ov{[V,U]}+2\Omega(V,U),\xi>=2<\Oa(V,\xi), U>\\
 &=&-2<\Oa(U,\xi),V>=-2<\ov{\Oa(U,\xi)},\ov V>\quad \blacksquare
\end{eqnarray*}

\begin{prop}
If $\theta\in C^\infty(M,\g)$ and $W\in\X M$ and $\R$ is the
curvature tensor of $M$ , then the curvature tensor , related to
$\n$, denoted by $\widehat \R$, satisfies the following
relations. The curvature tensor of $\n$, denoted by $\widehat
\R$,  satisfies the following relations.
 $\theta\in C^\infty(M,\g)$ and $W\in\X M$ and $\R$ is the curvature tensor of $M$.
\end{prop}
\begin{eqnarray}
\widehat \R(\xi,\eta)(\theta)&=&-{\te 1\over\te 4}[[\xi,\eta],\theta] \\
\widehat \R(\xi,\eta)(\ov W)&=&\ov{\Oa(\Oa(W,\eta),\xi)}-\ov{\Oa(\Oa(W,\xi),\eta)}
+\ov{\Oa(W,[\xi,\eta])} \\
\widehat \R(\xi,\ov V)(\theta)&=& \ov{\Oa(\Oa(V,\theta),\eta)}+{\te 1\over\te 2}\ov{\Oa(V,[\eta,\theta])}\\
\widehat \R(\xi,\ov V)(\ov W) &=& \ov{(\nabla_V\Oa)(W,\xi)}+\Omega(V,\Oa(W,\xi))+{\te 1\over\te 2}[\xi,\Omega(V,W)]\\
\widehat \R(\ov U,\ov V)(\theta) & =&-\ov{(\nabla_U\Oa)(V,\theta)}+\ov{(\nabla_V\Oa)(U,\theta)}\\
 && -\Omega(U,\Oa(V,\theta))+\Omega(V,\Oa(U,\theta))+[\Omega(U,V),\theta]\nonumber\\
\widehat \R(\ov U,\ov V)(\ov W)&=& \ov{\R(U,V)(W)}+(\nabla_U\Omega)(V,W)-(\nabla_V\Omega)(U,W)\\
&&-\ov{\Oa(U,\Omega(V,W))}+\ov{\Oa(V,\Omega(U,W))}\nonumber\\
&&+2\ov{\Oa(W,\Omega(U,V))}\nonumber
\end{eqnarray}
In the above relations, covariant derivation of $\Omega$ and
$\Oa$ is the combination of $\N$ and Levi-civita connection of
$M$.

{\bf proof:} Straightforward computations show these results. For
example we compute (18).
\begin{eqnarray*}
\widehat \R(\xi,\ov V)(\ov W) &=&\n_\xi\n_{\ov V}\ov W-\n_{\ov V}\n_\xi\ov W-\n_{[\xi,\ov V]}\ov W\\
&=&\n_\xi(\ov{\nabla_VW}+\Omega(V,W))-\n_{\ov V}(-\ov{\Oa(W,\xi)})-\n_{(-\N_V\xi)}\ov W\\
&=& -\ov{\Oa(\nabla_VW,\xi)}+{\te 1\over\te 2}[\xi,\Omega(V,W)]
+\ov{\nabla_V\Oa(W,\xi)}+\Omega(V,\Oa(W,\xi))\\
&&-\ov{\Oa(W,\N_V\xi)}\\
&=&\ov{(\nabla_V\Oa)(W,\xi)}+\Omega(V,\Oa(W,\xi))+ {\te 1\over\te 2}[\xi,\Omega(V,W)]\quad\blacksquare\\
\end{eqnarray*}
To compute Ricci curvature and scalar curvature of $TM^\g$, we
need to consider some orthonormal basis in $\g$ such as
$\{\theta_1,\cdots,\theta_k\}$ and  local orthonormal vector
fields on $M$ such as $W_1,\cdots,W_n$, in this case $\{
\theta_1,\cdots,\theta_k,\ov {W_1},\cdots,\ov {W_n} \}$ is an
orthonormal basis for $TM^\g$. Since $M$ is semi-Riemannian, then
$<W_j,W_j>=\pm 1$ .We set \mbox{$\hat j=<W_j,W_j>$}. Also denote
Killing form of $\g$ by $\B$ i.e. for $h,k\in\g,\ \B(h,k)=tr({\rm
ad}(h)\circ{\rm ad}(k))=-\sum_i<[h,\theta_i],[k,\theta_i]>$. In
the following we use inner products of tensors over $TM$ and
$\g$. Note that if $V$ and $W$ be some inner product spaces, we
can extend these inner products in the tensor spaces over $V$ and
$W$. For example, if $T,S:V\longrightarrow W$ be linear maps and
$\{e_1\ ,\ \cdots\ ,e_n\}$ be some orthonormal base Of $V$, and
$\hat j=<e_j\ ,\ e_j>$, then $<T,S>=\sum_j\hat j
<T(e_j),S(e_j)>$. Also, If $T,S:V\times V\longrightarrow W$ be
bilinear maps, then $<T,S>=\sum_{i,j}\hat j\hat i
<T(e_i,e_j),S(e_i,e_j)>$. These definitions do not depend on the
choice of the base.
\begin{prop}
The Ricci curvature tensor of $\n$, denoted by $\widehat {Ric}$,  satisfies the following relations.
\end{prop}
\begin{eqnarray}
\widehat {Ric}(\xi,\eta)&=&-{\te 1\over\te 4}\B(\xi,\eta)+<\Oa(.,\xi),\Oa(.,\eta)> \\
\widehat {Ric}(\xi,\ov V)&=&<{\rm div}\Oa(.,\xi),V> \\
\widehat {Ric}(\ov U,\ov V)&=& Ric(U,V)-2<\Omega(.,U),\Omega(.,V)>
\end{eqnarray}
{\bf proof:} Straightforward computations show these results.
\begin{eqnarray*}
\widehat {Ric}(\xi,\eta)&=&\sum_i<\widehat \R(\xi,\theta_i)(\theta_i),\eta>+\sum_j\hat j<\widehat
\R(\xi,\ov{W_j})(\ov{W_j}),\eta>\\
&=&\sum_i<-{\te 1\over\te 4}[[\xi,\theta_i],\theta_i],\eta>+\sum_j\hat j\left(
<\ov{(\nabla_{W_j}\Oa)(W_j,\xi)},\eta>\right.\\
&&\left. +<\Omega(W_j,\Oa(W_j,\xi)),\eta>+<{\te 1\over\te 2}[\xi,\Omega(W_j,W_j)],\eta>\right)\\
&=&{\te 1\over\te 4}\sum_i<[\xi,\theta_i],[\eta,\theta_i]>
+\sum_j\hat j<\Oa(W_j,\xi),\Oa(W_j,\eta)>\\
&=&-{\te 1\over\te 4}\B(\xi,\eta)+<\Oa(.,\xi),\Oa(.,\eta)>
\end{eqnarray*}
\begin{eqnarray*}
\widehat {Ric}(\xi,\ov V)&=& \sum_i<\widehat \R(\xi,\theta_i)(\theta_i),\ov V>+\sum_j\hat
j<\widehat \R(\xi,\ov{W_j})(\ov{W_j}),\ov V>\\
&=&\sum_i<-{\te 1\over\te 4}[[\xi,\theta_i],\theta_i],\ov V>+\sum_j\hat j\left(
<\ov{(\nabla_{W_j}\Oa)(W_j,\xi)},\ov V>\right.\\
&&\left. +<\Omega(W_j,\Oa(W_j,\xi)),\ov V>+<{\te 1\over\te 2}[\xi,\Omega(W_j,W_j)],\ov V>\right)\\
 &=&\sum_j\hat j<(\nabla_{W_j}\Oa)(W_j,\xi), V>=<{\rm div}\Oa(.,\xi),V>\\
 &&\\
\widehat {Ric}(\ov U,\ov V)&=&\sum_i<\widehat \R(\ov U,\theta_i)(\theta_i),\ov V>+\sum_j\hat
j<\widehat \R(\ov U,\ov{W_j})(\ov{W_j}),\ov V>\\
&=&\sum_i-< \ov{\Oa(\Oa(U,\theta_i),\theta_i)}+{\te 1\over\te
2}\ov{\Oa(U,[\theta_i,\theta_i])},\ov V>\\
&&\sum_j\hat j\left(\ov{\R(U,W_j)(W_j)},\ov V>-<\ov{\Oa(U,\Omega(W_j,W_j))},\ov V>\right. \\
&&\left. +<\ov{\Oa(W_j,\Omega(U,W_j))},\ov V>+2<\ov{\Oa(W_j,\Omega(U,W_j))},\ov V> \right)\\
&=&\sum_i<\Oa(U,\theta_i),\Oa(V,\theta_i)>+Ric(U,V)\\
&&-3\sum_j\hat j<\Omega(U,W_j),\Omega(V,W_j)>\\
\end{eqnarray*}
We can prove that these two sums are equal to $<\Omega(.,U),\Omega(.,V)>$.
\begin{eqnarray*}
\sum_i<\Oa(U,\theta_i),\Oa(V,\theta_i)>&=&\sum_i\sum_j\hat j<\Oa(U,\theta_i),W_j>
<\Oa(V,\theta_i),W_j>\\
&=&\sum_j\hat j\sum_i<\Omega(U,W_j),\theta_i)><\Omega(V,W_j),\theta_i)>\\
&=& \sum_j\hat j<\Omega(U,W_j),\Omega(V,W_j)>=<\Omega(.,U),\Omega(.,V)>
\end{eqnarray*}
So, (23) is proved.$\blacksquare$
\begin{prop}
If $R$ be the scalar curvature of $M$ then the scalar curvature
related to $\n$, denoted by $\widehat R$, satisfies the following
relation.
\end{prop}
\begin{equation}
\widehat R=R-<\Omega,\Omega>-{\te 1\over\te 4}tr(\B)
\end{equation}
 {\bf proof:}
\begin{eqnarray*}
\widehat R&=&\sum_i\widehat {Ric}(\theta_i,\theta_i)+\sum_j\widehat {Ric}(\ov {W_j},\ov {W_j})\\
&=&\sum_i\left(-{\te 1\over\te 4}\B(\theta_i,\theta_i)
+<\Oa(.,\theta_i),\Oa(.,\theta_i)>\right)\\
&&+\sum_j\left( Ric(W_j,W_j)-2<\Omega(.,W_j),\Omega(.,W_j)> \right)\\
&=&-{\te 1\over\te 4}tr(\B)+R+<\Oa,\Oa>-2<\Omega,\Omega>
\end{eqnarray*}
Since $<\Oa,\Oa>=<\Omega,\Omega>$, the proposition is proved. $\blacksquare$

\section{Application to general relativity and yang-mills theory}
In this section $M$ is a space-time, and $g$ and $\omega$ are same
as previous section. We can interpret $\omega$ as potential for
new relativistic forces in a similar way that yang-mills theory
describes. In this interpretation, geodesics of $TM^\g$ must show
path of point particles influenced by gravitation and new forces.
\begin{defn}
A smooth curve $\widehat\alpha:I\longrightarrow{TM}^\g$ is called a velocity-curve whenever there
exist a curve $\alpha:I\longrightarrow M$ such that $\rho(\widehat\alpha)=\alpha'$.
\end{defn}
In this case, there exists a curve $\xi:I\longrightarrow \g$ such
that $\widehat\alpha(t)=\ov{\alpha'(t)}+\xi(t)$. A smooth map
$\widehat\beta:I\longrightarrow{TM}^\g$ is called along the
velocity-curve $\widehat\alpha$, whenever
$\rho(\widehat\beta(t))\in T_{\alpha(t)}M$ i.e.
$\pi\circ\rho\circ\widehat\alpha=\pi\circ\rho\circ\widehat\beta$.
The covariant derivation of a map along a velocity-curve
$\widehat\alpha(t))$ is definable. A velocity-curve
$\widehat\alpha$ is called geodesic whenever
$\n_{\widehat\alpha}\widehat\alpha=0$.
\begin{prop}
A velocity-curve $\widehat\alpha(t)=\ov{\alpha'(t)}+\xi(t)$ is geodesic iff\\
\mbox{$\nabla_{\alpha'(t)}\alpha'(t)=2\Oa(\alpha'(t),\xi(t))$} and $\xi(t)$ is parallel along
$\alpha(t)$ with respect to $\N$.
\end{prop}
{\bf Proof:}
\begin{eqnarray*}
\n_{\widehat\alpha}\widehat\alpha &=&\n_{\ov{\alpha'(t)}+\xi(t)}\ov{\alpha'(t)}+\xi(t)\\
&=& \n_{\ov{\alpha'(t)}}\ov{\alpha'(t)}+\n_{\ov{\alpha'(t)}}\xi(t)+\n_{\xi(t)}\ov{\alpha'(t)}
+\n_{\xi(t)}\xi(t)\\
&=&\ov{\nabla_{\alpha'(t)}\alpha'(t)}-\ov{\Oa(\alpha'(t),\xi(t))}+\N_{\alpha'(t)}\xi(t)-\ov{\Oa(\alpha'(t),\xi(t))}
\end{eqnarray*}
Therefore, $\n_{\widehat\alpha}\widehat\alpha=0$ iff
$\nabla_{\alpha'(t)}\alpha'(t)=2\Oa(\alpha'(t),\xi(t))$ and
\mbox{$\N_{\alpha'(t)}\xi(t)=0$}. $\blacksquare$

In the case $\g=\Real$, $\xi$ represents ratio of charge to mass
and $2\Oa(\alpha'(t),\xi(t))$ represents electromagnetic force
exerted to the particle [3]. So we can consider elements of $\g$
as vector charges (divided by mass) and $2\Oa(\alpha'(t),\xi(t))$
can be considered as the force produced by these vector charges.

To find more detailed and explicit information of new forces and their analogy to
electromagnetism, it is better to write connection form $\omega$ and its curvature $\Omega$ with
respect to the base $\{ \theta_1,\cdots,\theta_k\}$.
\begin{eqnarray*}
\omega &=& \omega^i\theta_i\qquad \omega^i\in A^1(M)\ \ i=1,\cdots,k\\
\Omega &=& \Omega^i\theta_i\qquad \Omega^i\in A^2(M)\ \ i=1,\cdots,k
\end{eqnarray*}
Suppose $[\theta_i,\theta_j]=C_{ij}^l\theta_l$, since
$2\Omega(U,V)=d\omega(U,V)+[\omega(U),\omega(V)]$, by computation we find:
\begin{equation}
2\Omega^l=d\omega^l+C_{ij}^l\omega^i\wedge\omega^j
\end{equation}
Set $\Oa_i(U)=\Oa(U,\theta_i)$. $\Oa_i$ is the 1-1-form equivalent to 2-form $\Omega^i$.
\begin{cor}
Let $\widehat\alpha(t)=\ov{\alpha'(t)}+\xi(t)$ be a geodesic and $\xi(t)=\xi^i(t)\theta_i$, then
\end{cor}
\begin{eqnarray}
\nabla_{\alpha'(t)}\alpha'(t)&=&2\xi^l(t)\Oa_l(\alpha'(t))\\
{\xi^l}'(t)&=&C_{ij}^l\xi^i(t)\omega^j(\alpha'(t))
\end{eqnarray}
We can interpret $\xi^l$ as $l-$th charge and $\Oa_l$ as $l-$th
electromagnetism field and $2\xi^l(t)\Oa_l(\alpha'(t))$ as sum of
the the forces exerted by these fields. Of course these forces
are not independent  and (25) shows these forces are dependent to
each other and are components of a more general force.

Set ${\bf B}_{ij}={\bf B}(\theta_i,\theta_j)$. $\widehat{Ric}$ and $\widehat R$ with respect to
the base $\{ \theta_1,\cdots,\theta_k\}$, can be written as follows:
\begin{eqnarray}
\widehat {Ric}(\theta_i,\theta_j)&=&-{\te 1\over\te 4}\B_{ij}+<\Oa_i,\Oa_j> \\
\widehat {Ric}(\theta_i,\ov V)&=&<{\rm div}\Oa_i,V> \\
\widehat {Ric}(\ov U,\ov V)&=& Ric(U,V)-2\sum_i<\Oa_i(U),\Oa_i(V)>\\
\widehat R&=&R-\sum_i<\Oa_i,\Oa_i>-{\te 1\over\te 4}\sum_i\B_{ii}
\end{eqnarray}
 In the case of $\g=\Real$, energy-momentum tensor of
electromagnetism forces in a suitable system of measurement,($c=1\ ,\ G=1\ ,\ \epsilon_0={\te 1\over\te 16\pi}$)[3], is
defined as follows.
\begin{equation}
 {\bf T}^{elec}={\textstyle 1\over\textstyle  4\pi}(<\Oa(.),\Oa(.)>-{\textstyle
1\over\textstyle 4}<\Oa,\Oa>{\bf g})
\end{equation}
${\bf g}$ is the tensor metric of $M$. We can extend this
definition and define $i-$th  energy-momentum tensor of $\omega$
by
\begin{equation}
{\bf T}^\omega_i={\textstyle 1\over\textstyle
4\pi}(<\Oa_i(.),\Oa_i(.)>-{\textstyle 1\over\textstyle
4}<\Oa_i,\Oa_i>{\bf g})
\end{equation}
Define the whole energy-momentum tensor of $\omega$ by
$T^\omega=\sum_i T^\omega_i$. This definition dose not depend on
the choice of $\theta_i$, in fact:
\begin{equation}
{\bf T}^\omega(U,V)={\textstyle 1\over\textstyle
4\pi}(<\Oa(U,.),\Oa(V,.)>-{\textstyle 1\over\textstyle
4}<\Oa,\Oa><U,V>)
\end{equation}
Denote the meter of ${TM}^\g$ by $\widehat{\bf g}$. To construct
a suitable field equation that produces Einstein field equation,
we should imitate Einstein field equation in the context of this
algebroid bundle. So by analogy, we can consider
 \mbox{$\widehat G=\widehat{Ric}-{\te 1\over\te 2}\widehat R\widehat{\bf g}$}
 as the extended Einstein tensor. This tensor in block form looks like the following matrix.
\[
\left( \begin{array}{c|c} Ric -{1\over 2}R{\bf g}+{1\over 8}{\rm tr}(\B){\bf g}-8\pi T^\omega& {\rm div}\Oa \\
\hline ^t{\rm div}\Oa &  \lambda_{ij}
\end{array}\right)
\]
\[
\lambda_{ij}=-{\te 1\over\te 4}\B_{ij}+<\Oa_i,\Oa_j>-{1\over
2}\widehat R\delta_{ij}
\]
Note that $\widehat R=R-<\Omega,\Omega>-{\te 1\over\te 4}{\rm
tr}(\B)$. Now, we can construct vacuum field equation as follows.
\begin{equation}
\widehat{Ric}-{\te 1\over\te 2}\widehat R\widehat{\bf g}=0
\end{equation}
This equation yields $\widehat R=0$ and is equivalent to
$\widehat{Ric}=0$ and has the following consequences.
\begin{eqnarray}
Ric -{1\over 2}R{\bf g}+{1\over 8}{\rm tr}(\B){\bf g}&=&8\pi T^\omega\\
{\rm div}\Oa_i &=&0\\
<\Oa_i,\Oa_j> &=&{\te 1\over\te 4}\B_{ij}
\end{eqnarray}
Note that in $\lambda_{ij}$, we have $\widehat R=0$, so
$\lambda_{ij}=0$ yields (38). Two of these equations are Einstein
and yang-mills field equations in vacuum and we find a third new
equation, (38), that may have new results. Moreover, Einstein
field equation naturally yields  a cosmological constant that
depends on inner product of $\g$ and by re-scaling  can be adapted
its value to experimental data.

Particles are modeled by representations of the Lie algebra $\g$.
Suppose  a representation of $\g$ on some inner product vector
spaces $W$ that $\g$ acts on $W$ anti-symmetrically. Any $X\in
C^\infty (M,W)$ is called a particle field. We can consider
$\rho=<X,X>$ as density of this particle field. Charge density of
a particle field can be considered as a smooth function
$\eta:M\longrightarrow\g$.

In order to construct the field equation including matter, we
should extend the concept of energy-momentum tensor of matter,
and we do this the same as [3]. Let $T^{\rm mass}$ be the ordinary
energy-momentum tensor of the particle field. For every observer
$Z$, $T^{\rm mass}(Z,Z)$ is the energy of the particle measured
by $Z$[8]. We can define current of this particle field to be
$J\in A^1(M,\g)$ such that for every observer $Z$, $J(Z)$ is the
vector charge of the particle field measured by $Z$.

$T^{\rm mass}$ and $J$ are parts of the extended energy-momentum
tensor which is denoted by $\widehat T$. In fact for $U,V\in\X M$
and $\xi\in C^\infty (M, \g)$, we define:
\[ \widehat T(\bar U,\bar V)=T^{\rm mass}(U,V)\quad , \quad
\widehat T(\bar U,\xi)=<J(U),\xi> \]
 To complete the
construction of $\widehat T$, similar to [3], we need a symmetric
2-tensor on $\g$. It seems we should consider ${\te
1\over\te\rho}\eta\otimes\eta$ for this tensor. Because, in the
case $\g=I\!\! R$ in [3] we have good reason for it and it is
very natural. Of course, only results of this choice and
experience can show that this choice is true or not. So we
propose, $\widehat T(\xi_1,\xi_2)={1\over\te
\rho}<\xi_1,\eta><\xi_2,\eta>$, and $\widehat T$ be defined as
follows:
\begin{equation}
\widehat T=\left(\begin{array}{c|c}
T^{\rm mass} & ^t\! J\\ \hline J & {\te 1\over\te\rho}\eta\otimes\eta
\end{array}\right)
\end{equation}
 Now, we can write  the
 Einstein field equation in this structure as following.
\begin{equation}
\widehat{Ric}-{\te 1\over\te 2}\widehat R\widehat{\bf g}=8\pi\widehat T
\end{equation}
This equation contains three following equations.
\begin{eqnarray}
Ric -{1\over 2}R{\bf g}+{1\over 8}{\rm tr}(\B){\bf g}&=&8\pi (T^\omega+T^{\rm mass})\\
^t{\rm div}\Omega &=&8\pi J\\
\lambda_{ij}&=&8\pi{\te \eta_i\eta_j\over\te\rho}
\end{eqnarray}
The first two equations are Einstein and yang-mills field
equations and third equation is new and may have new results.
Equation (40) makes Einstein and yang-mills theory into a unified
theory in the context of lie algebroid structures.

\section{conclusion}
These constructions retrieve yang-mills theory in the context of
lie algebroid structures and they need no principle bundle and
principle connection. This theory is more simple and does not make
any extra dimension in space-time, instead it enriches tangent
bundle by a lie algebra and makes a lie algebroid and replaces
tangent bundle by this lie algebroid.

Of course, this theory does not contain quantum effects and internal structures of particles. This
theory must be improved such that internal structures of particles determine density and vector
charge density naturally.


\begin{thebibliography}{12}
\bibitem{}D. Bleeker, Gauge Theory and Variational Principles, addison-wesley (1981)
\bibitem{}M. Boucetta, Riemannian Geometry of Lie algebroids, arxiv:0806.3522v2 (2008)
\bibitem{}N. Elyasi, N. Boroojerdian, Affine metrics and algebroid structures: Application to general relativity and
unification, arxiv:1108.2843v1(2011)
\bibitem{}K. Grabowska, J.Grabowski, P.Urbanski, Geometrical Mechanics on algebroids, int. J. Geom. Method. Phys. 3(2006)
\bibitem{} Myroon, W. Evans, Generally covariant Unified Field Theory, abramis (2005)
\bibitem{} J. M. Overdain: Kaluza-Klien gravity:arxiv:gr-qc/9805018v1(1998)
\bibitem{} W. A. Poor, Differential Geometric Structures, McGraw-Hill,
1981.
\bibitem{} R.K. Sachs and  H. Wu,  General Relativity for Mathematicians, Springer-verlag, New York, 1977.
\bibitem{} Paul S Wesson: Space time matter: Modern kaluza klien theory:word scientific.



\end{thebibliography}
\end{document}